# TrackRAD2025 challenge dataset: Real-time tumor tracking for MRI-guided radiotherapy

Yiling Wang[1+], Elia Lombardo[2+], Adrian Thummerer[2], Tom Blöcker[2], Yu Fan[1], Yue Zhao[1], Christianna Iris Papadopoulou[2], Coen Hurkmans[3, 17], Rob H.N. Tijssen[3, 18], Pia A.W. Görts[3, 18], Shyama U. Tetar[3], Davide Cusumano[4], Martijn PW Intven[5], Pim Borman[5], Marco Riboldi[7], Denis Dudáš[8], Hilary Byrne[9], Lorenzo Placidi[10], Marco Fusella[11], Michael Jameson[12], Miguel Palacios[13], Paul Cobussen[13], Tobias Finazzi[13], Cornelis J.A. Haasbeek[13], Paul Keall[14], Christopher Kurz[2], Guillaume Landry*[2, 15, 16], Matteo Maspero[5,6]

1 Department of Radiation Oncology, Radiation Oncology Key Laboratory of Sichuan Province, Sichuan Clinical Research Center for Cancer, Sichuan Cancer Hospital & Institute, Sichuan Cancer Center, University of Electronic Science and Technology of China, Chengdu, China
2 Department of Radiation Oncology, LMU University Hospital, LMU Munich, Munich, Germany
3 Department of Radiation Oncology, Catharina Hospital, Eindhoven, The Netherlands
4 Medical Physics Unit, Mater Olbia Hospital, Olbia, Italy
5 Department of Radiotherapy, University Medical Center Utrecht, Utrecht, The Netherlands;
6 Computational Imaging Group for MR Diagnostics & Therapy, University Medical Center Utrecht, Utrecht, The Netherlands;
7 Department of Medical Physics, Ludwig-Maximilians-Universität München, Garching, Germany
8 Faculty of Nuclear Sciences and Physical Engineering, Czech Technical University in Prague.
9 Faculty of Medicine and Health, Image X Institute, University of Sydney, Darlington, New South Wales, Australia
10 Fondazione Policlinico Universitario Agostino Gemelli, IRCCS, Department of Diagnostic Imaging, Oncological Radiotherapy and Hematology, Rome, Italy
11 Department of Radiation Oncology, Abano Terme Hospital, Abano Terme Veneto, Italy
12 GenesisCare, St Vincent's Hospital, Sydney, Australia
13 Department of Radiation Oncology, Amsterdam UMC location Vrije Universiteit Amsterdam, Amsterdam, The Netherlands
14 Image X Institute, Faculty of Medicine and Health, University of Sydney, Sydney, Australia.
15 German Cancer Consortium (DKTK), Partner Site Munich, a Partnership between DKFZ and LMU University Hospital, Munich, Germany
16 Bavarian Cancer Research Center (BZKF), Munich, Germany
17 Department of Electrical Engineering and Department of Applied Physics and Science Education, Technical University Eindhoven, The Netherlands
18 Department of Biomedical Engineering, Technical University Eindhoven, Eindhoven, the Netherlands

+ These authors contributed equally.
* Corresponding author e-mail: guillaume.landry@med.uni-muenchen.de

**Abstract**

**Purpose:**

Magnetic resonance imaging (MRI) to visualize anatomical motion is becoming increasingly important when treating cancer patients with radiotherapy. Hybrid MRI-linear accelerator (MRI-linac) systems allow real-time motion management during irradiation. This paper presents a multi-institutional real-time MRI time series dataset from different MRI-linac vendors. The dataset is designed to support developing and evaluating real-time tumor localization (tracking) algorithms for MRI-guided radiotherapy within the TrackRAD2025 challenge (https://trackrad2025.grand-challenge.org/).

**Acquisition and validation methods:**

The dataset consists of sagittal 2D cine MRIs (20-20543 frames per scan) in 585 patients from six centers (3 Dutch, 1 German, 1 Australian, and 1 Chinese). Tumors in the thorax, abdomen, and pelvis acquired on two commercially available MRI-linacs (0.35 T and 1.5 T) were included. For 108 cases, irradiation targets or tracking surrogates were manually segmented on each temporal frame. The dataset was randomly split into a public training set of 527 cases (477 unlabeled and 50 labeled) and a private testing set of 58 cases (all labeled).

**Data Format and Usage Notes:**

The data is publicly available under the TrackRAD2025 collection: https://doi.org/10.57967/hf/4539. Both the images and segmentations for each patient are available in metadata format.

**Potential Applications:**

This novel clinical dataset will enable the development and evaluation of real-time tumor localization algorithms for MRI-guided radiotherapy. By enabling more accurate motion management and adaptive treatment strategies, this dataset has the potential to advance the field of radiotherapy significantly.

## 1. Introduction

Over the last decades, magnetic resonance imaging (MRI) has proven its added value in tumor and organs-at-risk delineation thanks to its superb soft-tissue contrast [1]. MRI to visualize and characterize motion is becoming increasingly important in treating cancer patients with radiotherapy. Respiratory motion can lead to large tumor displacements and, therefore, a decreased dose to the target and an increased dose to healthy tissue if not properly managed [2-4]. The recent development of MRI-guided radiotherapy (MRIgRT) using hybrid MRI-linear accelerator (MRI-linac) systems [5, 6] offers the possibility to adapt to changes in tumor position during treatment in real-time. Clinically, MRI-linacs support intra-fractional motion management through a cine MRI-based gating approach [7-11], where either a single 2D or interleaved orthogonal 2D cine MRI planes are continuously acquired throughout the irradiation session. Recent research has explored multi-leaf-collimator (MLC)-tracking as a more time-efficient yet dosimetrically equivalent alternative to gating [12-14].

A key challenge in motion management workflows, particularly for MLC-tracking, is the accurate localization of the irradiation target (or a surrogate structure) on 2D cine MRIs acquired at 1-8 Hz during the irradiation [15]. Target or surrogate localization must occur with high accuracy and robustness to ensure the sparing of critical organs. Currently, clinically available solutions rely on conventional deformable image registration (DIR) to propagate contours from a labeled reference frame, or template matching. Still, these methods struggle with large, non-rigid motion [16-18]. Recently, artificial intelligence (AI) algorithms, especially deep learning models, have shown promise for target tracking in MRIgRT [15, 19] as they offer fast execution and may outperform conventional methods [20-22].

However, to date, no public datasets or challenges have been designed to benchmark different MRIgRT tumor-tracking approaches. A recent review on AI-enabled real-time motion management in MRIgRT advocated for public challenges to provide shared data and standard end-to-end tests for real-time adaptive MRIgRT [15]. In MRIgRT, real-time imaging relies on cine-MRI with frame rates of up to 8 Hz. Due to the high frame rate and corresponding large number of frames, labeling 2D cine MRIs is time-consuming. A comprehensive database of labeled and unlabeled data would be highly valuable to the community, facilitating the development of various supervised and unsupervised training strategies.

This dataset is released in support of the TrackRAD2025 challenge (https://trackrad2025.grand-challenge.org/). The objective of TrackRAD2025 is to provide the first public multi-institutional dataset and evaluation platform to compare the latest developments in online time-resolved 2D cine-MRI-based tracking methods. TrackRAD will provide a template tumor or surrogate segmentation on the first frame and the remaining 2D cine-MRI sequence requiring real-time segmentation. Algorithms will be evaluated based on their ability to reproduce ground truth segmentation labels. We expect the multi-vendor, multi-center dataset of 2D cine MRIs to allow benchmarking competitive real-time tumor localization algorithms based on a unified platform, thus coming closer to the ultimate goal of high-precision MRIgRT.

## 2. Acquisition and validation methods

### 2.1 Overview of the dataset

The dataset comprises 2D sagittal cine MRI sequences acquired from 585 patients across six international centers (indicated with the letters A-F) using 0.35 T MRIdian (ViewRay Inc., Oakwood, USA) and 1.5 T Unity (Elekta AB, Stockholm, Sweden) MRI-linac systems. Participating institutions included the Amsterdam University Medical Center, Sichuan Cancer Hospital Chengdu, Catharina Hospital Eindhoven, LMU University Hospital Munich, GenesisCare Sydney, and University Medical Center Utrecht. All acquisitions occurred during radiotherapy sessions with patients in treatment positions, covering thoracic (32%), abdominal (41%), and pelvic (27%) anatomical regions. Ethical approval for the use of patient data was obtained from the respective institutional review boards (IRBs) at each participating center. Patient consent was obtained according to local regulations. Each center was assigned randomly with a unique identifier (A to F) to maintain anonymity. All 0.35 T data were acquired before upgrades to the latest A3i version. Data at a sampling rate of 8 Hz were from MRIdian devices where the SmartVision upgrade was installed. One of the centers installed Elekta comprehensive motion management (CMM) and additionally provided data with an updated cine MRI sequence for gating at 1.5 T MRI-linacs. The CMM data from this center is indicated explicitly with the letter X.

At the three centers using the 0.35 T MRI-linac, the 2D cine MRIs were acquired in sagittal orientation with the patient in treatment position in the MRI-linac bore. During treatment simulation or delivery, the patients performed breath holds to increase the duty cycle of the gated radiation delivery. Breath-holds lasted 20-30 seconds, followed by free breathing periods of variable lengths. The used sequence was a 2D-balanced steady-state free precession (bSSFP, named TrueFISP on the scanner) with Cartesian acquisition at 4.0 Hz sampling frequency, relaxation time (TR) = 2.0-2.4 ms, echo time (TE) = 0.9-1.1 ms, flip angle (FA) = 60°, bandwidth = 1000-1351 Hz/px, or a radial sequence with 8.0 Hz sampling frequency, TR=2.7-3.4 ms, TE=1.4-1.7 ms, FA=95°, bandwidth = 558-890 Hz/px, slice thickness = 5, 7 or 10 mm and pixel spacing = 2.4×2.4 mm or 3.5×3.5 mm. Due to machine design, the image quality is degraded when the gantry of the 0.35 T MRI-linac moves. The labeled data was chosen in such a way as to avoid any image degradation, while the unlabeled data still contains degraded frames.

At the three centers using the 1.5 T MRI-linac, the 2D cine-MRIs were acquired in interleaved sagittal and coronal or interleaved sagittal, coronal, and axial orientations with the patient in treatment position in the bore. For the challenge, only sagittal planes have been selected, and the reported sampling frequency refers to sampling in that plane only. During treatment simulation or delivery, some patients performed breath holds to increase the duty cycle of the gated radiation delivery, while others breathed freely. Breath-holds lasted 20-30 seconds, followed by free breathing periods of variable lengths. The used sequence was also bSSFP (named bFFE on the scanner) at frequencies of 1.3 Hz to 3.5 Hz with TR = 2.7-4.2 ms, TE = 1.4-2.1 ms, FA = 40-50°, bandwidth = 1078 Hz/px, slice thicknesses = 5, 7 or 8 mm and pixel spacing = 1.0×1.0 mm to 1.7×1.7 mm in the sagittal orientation.

Table 1 summarizes the demographic and image acquisition parameters for each center. Treatments were occasionally interrupted due to, for instance, re-positioning of the patient or trying different MRI acquisition settings. For 1.5 T MRI-linac data, these interruptions are not flagged anywhere, and all scans of the same fraction get exported as a single series of frames. By visual inspection, we avoided having temporal discontinuities in the labeled data. However, these are present in the unlabeled dataset.

The dataset consists of over 2.8 million unlabeled 2D sagittal cine MRI frames from 477 patients and over 10,000 manually labeled 2D sagittal cine MRI frames from another 108 patients. Specifically, there are 253 unlabeled cases from 0.35 T and 224 unlabeled cases from 1.5 T MRI-linacs with approximately 15% abdominal nodes or kidney, 25% liver, 30% lung, 10% pancreas, 10% prostate, 5% gynecological and 5% rectal cancer patient cases. In addition, there are 52 labeled cases from 0.35 T and 56 labeled from 1.5 T MRI-linacs with approximately 15% abdominal nodes or kidney, 20% liver, 30% lung, 15% pancreas, and 10% prostate cases. We dropped the first five frames (0.35 T) or the first three (1.5 T) from each time series to ensure a steady state during the MRI. For the labeled dataset, in case the resulting first frame did not contain the tracking target (due to out-of-plane motion), we also dropped consecutive frames until the target was present. In most cases, the tracking target was also the irradiation target, i.e., the gross tumor volume. In some cases, the visibility/contrast of the tumor was considered too low clinically, and a surrogate structure was tracked. A concrete example is the tracking of the entire liver instead of a low-contrast liver lesion. In such cases, we manually segmented the surrogate structure instead of the tumor, following the clinical choices.

**Table 1: Image acquisition parameters for each center. The software versions refer to the MRI system of the MRI-linacs.**

| **Center** | **Magnetic Field** [T] | **Mean Age** [years] | **Male / Female** [%] | **Sampling frequency** [Hz] | **Slice thickness** [mm] | **Pixel spacing** [mm] | **TE** [ms] | **TR** [ms] | **MRI-linac Software version** |
|---|---|---|---|---|---|---|---|---|---|
| **A** | 0.35 | 65.6 | 59/41 | 4.0 or 8.0 | 5, 7 or 10 | 2.4×2.4 or 3.5×3.5 | 0.9-1.1 or 1.4-1.7 | 2.0-2.4 or 2.7-4.2 | Syngo MR B19 |
| **B** | 1.5 | 66.7 | 64/36 | 1.3 -3.5 | 5, 7 or 8 | 1.0×1.0 to 1.7×1.7 | 1.4-2.1 | 2.7-4.2 | Marlin 5.7 |
| **C** | 1.5 | 59.7 | 70/30 | 1.7 | 5 | 1.2×1.2 or 1.3×1.3 | 1.4-2.1 | 2.7-4.2 | Marlin 6.1 |
| **D** | 0.35 | 68.0 | 55/45 | 4.0 | 5, 7 or 10 | 2.4×2.4 or 3.5×3.5 | 0.9-1.1 or 1.4-1.7 | 2.0-2.4 or 2.7-4.2 | Syngo MR B19 |
| **E** | 0.35 | 72.9 | 65/35 | 4.0 or 8.0 | 5, 7 or 10 | 2.4×2.4 or 3.5×3.5 | 0.9-1.1 or 1.4-1.7 | 2.0-2.4 or 2.7-4.2 | Syngo MR B19 |
| **F** | 1.5 | 68.4 | 80/20 | 1.3-3.5 | 5, 7 or 8 | 1.0×1.0 to 1.7×1.7 | 1.4-2.1 | 2.7-4.2 | Marlin 5.8.1 |
| **X** | 1.5 | 63.7 | 83/17 | 3.0 | 5, 7 or 8 | 1.0×1.0 to 1.7×1.7 | 1.4-2.1 | 2.7-4.2 | Marlin 5.8.1 |

## 2.2 Labeling

Human observers manually performed the ground truth annotations for the labeled dataset for the training and testing sets. Specifically, for dataset A, two observers (a medical student and a dentistry student) labeled the cine MRI frames using a labeling tool developed specifically for the TrackRAD2025 challenge (https://github.com/LMUK-RADONC-PHYS-RES/contouring-tool). For dataset B, a medical physics researcher (assistant professor) with over 10 years of experience in radiotherapy used the same in-house labeling tool to delineate the frames. Two radiation oncologists independently labeled the cine MRI frames for dataset C using ITK-SNAP [23]. For dataset D, four radiation oncologists and one medical physicist independently labeled the cine MRI frames using software provided by the 0.35 T MRI-linac vendor. Centers E and F only provided unlabeled datasets. Since center X was one of the centers A to D, the labeling approach was the same as that center. For all labeled data, a medical physics doctoral student with four years of experience in tumor tracking reviewed and, if necessary, corrected all labels using the in-house tool. Typical errors corrected included erroneous clicks generating isolated small regions in the segmentation or inconsistent inclusion of structures throughout a cine MRI.

Inter-observer annotation variability remains an issue when evaluating target tracking accuracy. For cases with multiple annotators, the simultaneous truth and performance level estimation (STAPLE) algorithm was used to generate a single ground truth [24]. For a given frame, the TrackRAD2025 geometric evaluation metrics were calculated as the median over the five observers vs the STAPLE ground truth. For a given metric, the resulting medians per frame were averaged over all frames of a given case. We took the median of the average metric per case over all cases. In this procedure, we used the median when the data was not normally distributed (observers and cases) and the mean when it was (over the frames of one scan). We applied this procedure to the five observers of center D. The results showed a Dice similarity coefficient (DSC) of 0.94 ± 0.07, indicating high structure overlap [25]. The 95$^{th}$ percentile Hausdorff distance (HD95) was 4.2 ± 2.8 mm, representing the 95$^{th}$ percentile of all distances from a boundary point to the closest point on the other boundary [26]. The mean average surface distance (MASD) was 1.0 ± 1.7 mm, measuring the mean shortest distances between corresponding boundary points [27]. Lastly, the mean Euclidean center distance (CD) was 1.2 ± 0.9 mm, quantifying the Euclidean distance between the object's reference and predicted center points [28].

### 2.3 Image export, preprocessing, and data split

All images were acquired using the clinically adopted imaging protocols of the respective centers for each anatomical site and reflect typical images found in daily clinical routines. The cine-MRI sequences used at the 0.35 T and 1.5 T MRI-linacs are standardized, ensuring the data's uniformity for a given field strength. The data was available for export in DICOM format for one 1.5 T center, while it was only exportable in proprietary binary formats for the others. To convert the cine MRI frames from proprietary formats to MetaImage (MHA) format, the centers with MRIdian MRI-linac systems used scripts provided by the vendor. In contrast, the centers with Unity MRI-linac systems used either a tool provided by the vendor or an in-house pipeline first to convert the binary files into metadata files and then classify the metadata files into sagittal, coronal, and axial using a machine learning algorithm that was trained for this purpose. Only the sagittal frames were then selected for the challenge. The converted MHA data was then resampled to 1×1 mm using bilinear interpolation, stored as a 16-bit unsigned integer, and then compressed using the SimpleITK library (https://simpleitk.org/). The annotation occurred on the resampled images, and the segmentations were stored as 8-bit unsigned integers. More information and the code for the in-house format conversion pipeline for both the MRIdian and Unity cine MRI data can be found on GitHub (https://github.com/LMUK-RADONC-PHYS-RES/trackrad2025/tree/main/scripts/proprietary_format_conversion).

The training set comprising 477 unlabeled cases plus 50 labeled cases was publicly released on HuggingFace (https://doi.org/10.57967/hf/4539). The remaining 58 labeled cases building the preliminary test (8 cases) and final testing set (50 cases) are only accessible for evaluation via submission to the challenge. The preliminary testing phase allows each team to familiarize themselves with the submission system. For training, centers A, B, and C provided unlabeled and manually labeled data, whereas centers E and F provided solely unlabeled data. Centers A, B, C, D, and X provided manually labeled data for preliminary and final testing.

Table 2 summarizes the number of training and testing cases of each center per anatomical site and the distribution of the labeled and unlabeled cases. For the labeled dataset, approximately 20% of the patients performed breath-holds during acquisitions with 0.35 T MRI-linacs, while 30% did so during acquisitions with 1.5 T MRI-linacs. Figure 1 displays unlabeled cine MRI frames with tumors from different anatomical sites and MRI-linacs. Figure 2 displays a time series of labeled cine MRI frames from two patients imaged with a 0.35 T and a 1.5 MRI-linac, respectively. Figure 3 displays the mean and variability in target sizes for the 50 labeled cases in the training set, where the sizes ranged from 0.53 to 212.8 $cm^2$ with a medium value and inter-quartile range (IQR) of 9.8 $cm^2$ and 18.8 $cm^2$, respectively. To display the variability in target sizes, we normalized each target size by its median value. Therefore, the line inside each box is always 1 in Figure 3(b). The DSC and the CD between the first and the consecutive frames of the 50 labeled training cases are shown in Figure 4. For DSC, the value ranged from 0.0 to 1.0 with a medium value of 0.85 and IQR of 0.22 across all the patients. For CD, the value ranged from 0.0 to 69.9 mm with a medium value of 3.1 mm and IQR of 7.1 mm across all the patients.

**Table 2: Number of training and testing cases provided by each center per anatomical site. L and U stand for labeled and unlabeled cases, respectively.**

| **Center** | **Training (L & U)** | | | **Testing (L)** | | | | | |
|---|---|---|---|---|---|---|---|---|---|
| | | | | **Preliminary** | | | **Final** | | |
| | Thorax | Abdomen | Pelvis | Thorax | Abdomen | Pelvis | Thorax | Abdomen | Pelvis |
| **A** | 8 & 95 | 15 & 99 | 2 & 25 | 1 | 1 | 0 | 2 | 2 | 1 |
| **B** | 2& 21 | 13 & 24 | 0 & 18 | 1 | 2 | 0 | 4 | 4 | 0 |
| **C** | 5 & 17 | 1 & 24 | 4 & 19 | 0 | 1 | 2 | 4 | 3 | 4 |
| **D** | 0 & 0 | 0 & 0 | 0 & 0 | 0 | 0 | 0 | 5 | 15 | 0 |
| **E** | 0 & 13 | 0 & 15 | 0 & 6 | 0 | 0 | 0 | 0 | 0 | 0 |
| **F** | 0 & 1 | 0 & 41 | 0 & 59 | 0 | 0 | 0 | 0 | 0 | 0 |
| **X** | 0 & 0 | 0 & 0 | 0 & 0 | 0 | 0 | 0 | 0 | 6 | 0 |

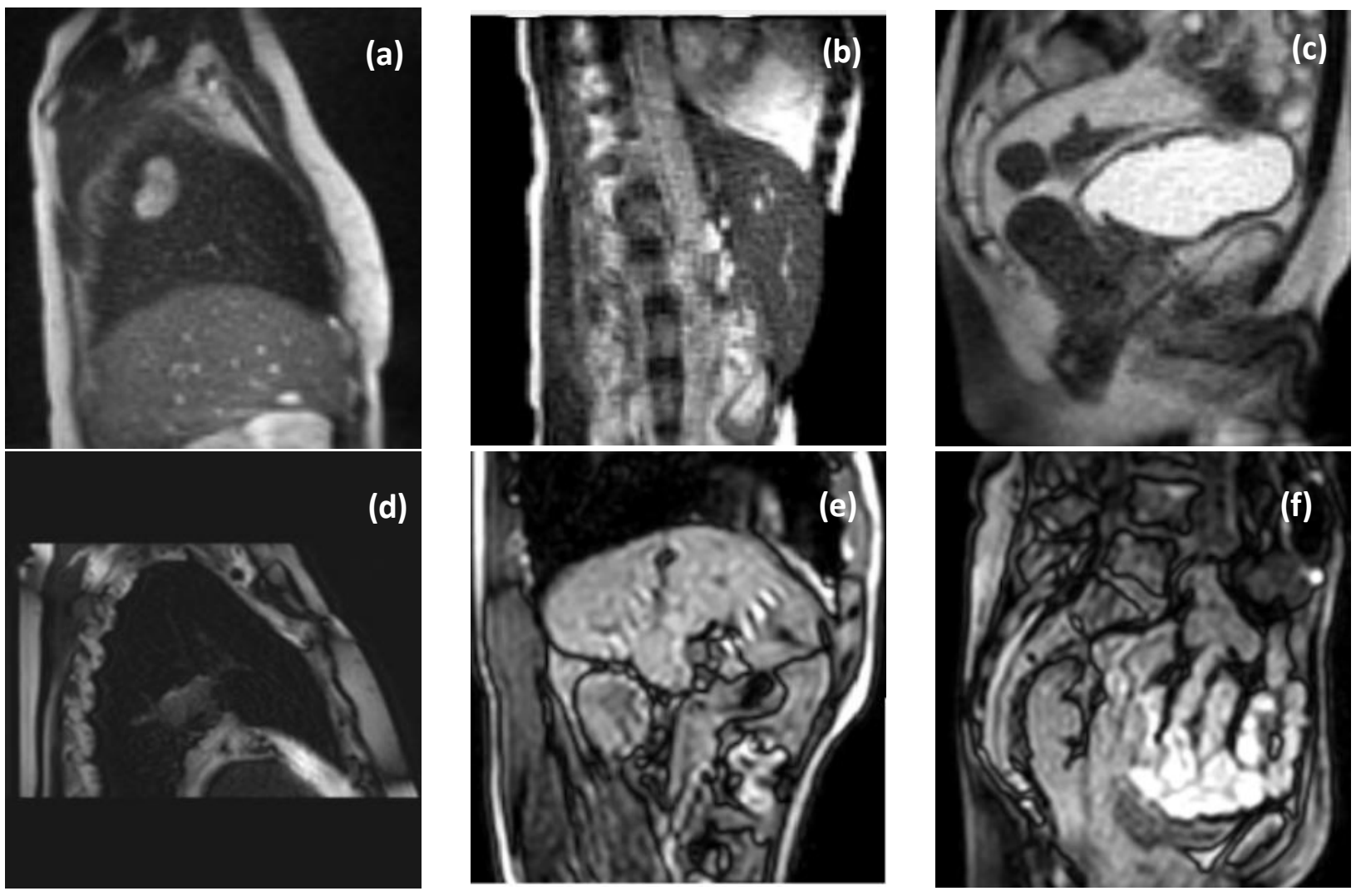

**Figure 1: Unlabeled frames from different anatomies and MRI-linacs. (a) to (c) show frames from a 0.35 T MRI-linac and (d) to (f) show frames from a 1.5 T MRI-linac, respectively. (a) and (d), (b) and (e), (c), and (f) show thoracic, abdominal, and pelvic cases, respectively.**

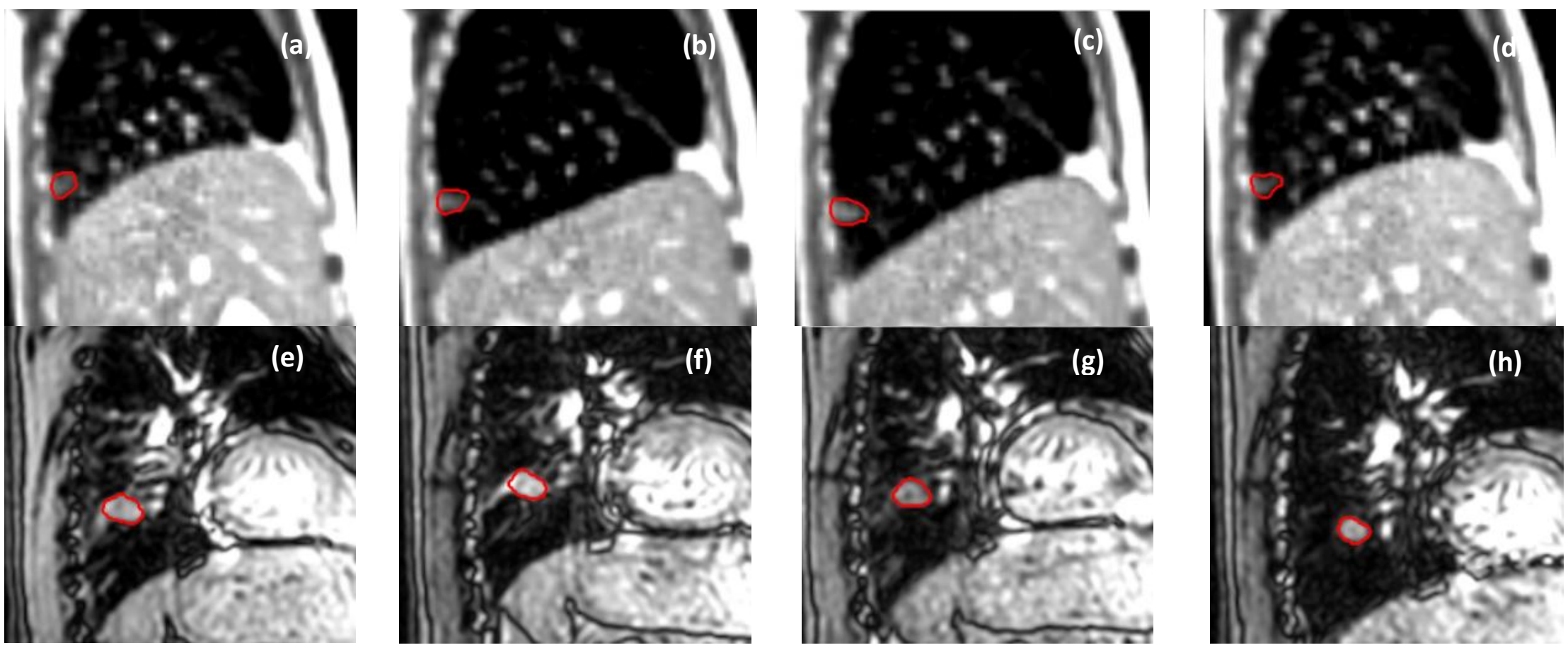

**Figure 2: Time sequence of labeled frames. (a) to (d) show frames from a 0.35 T MRI-linac and (e) to (h) show**

**frames from a 1.5 T MRI-linac, respectively. The label for each frame is shown in red.**

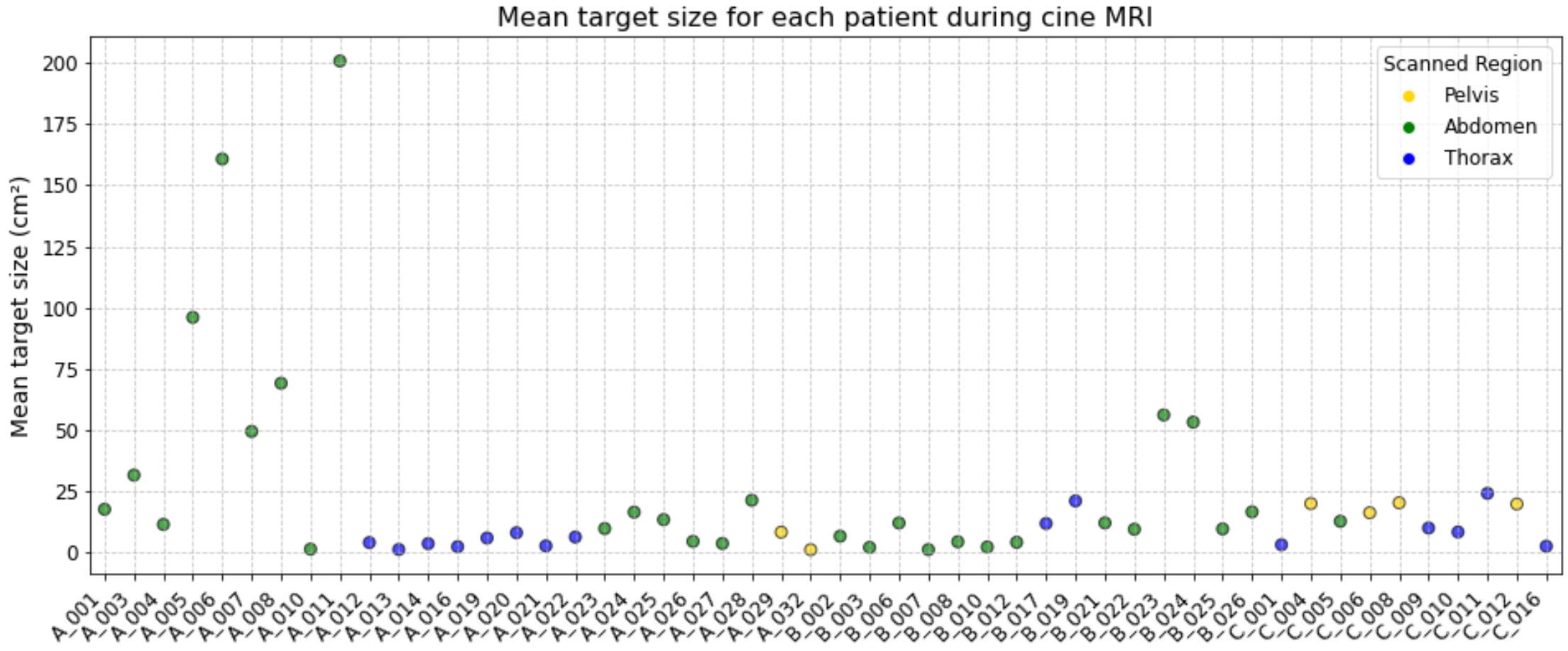

(a)

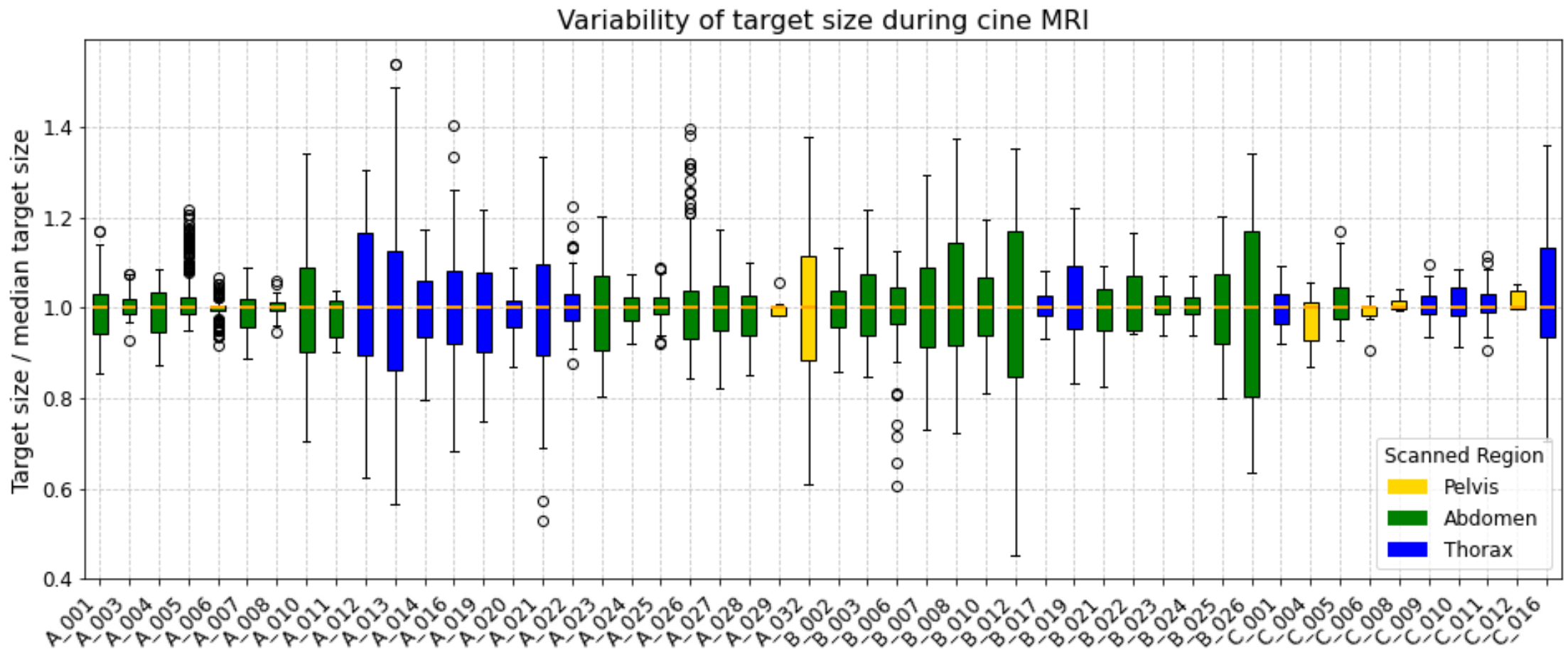

(b)

**Figure 3: The mean (a) and variability (b) in target sizes for the 50 labeled cases in the training set. The lower and upper box boundaries represent the 25th and 75th percentiles, respectively, the line inside the box represents the median, the whiskers extend to the 10th and 90th percentiles and circles represent data falling outside 10th and 90th percentiles.**

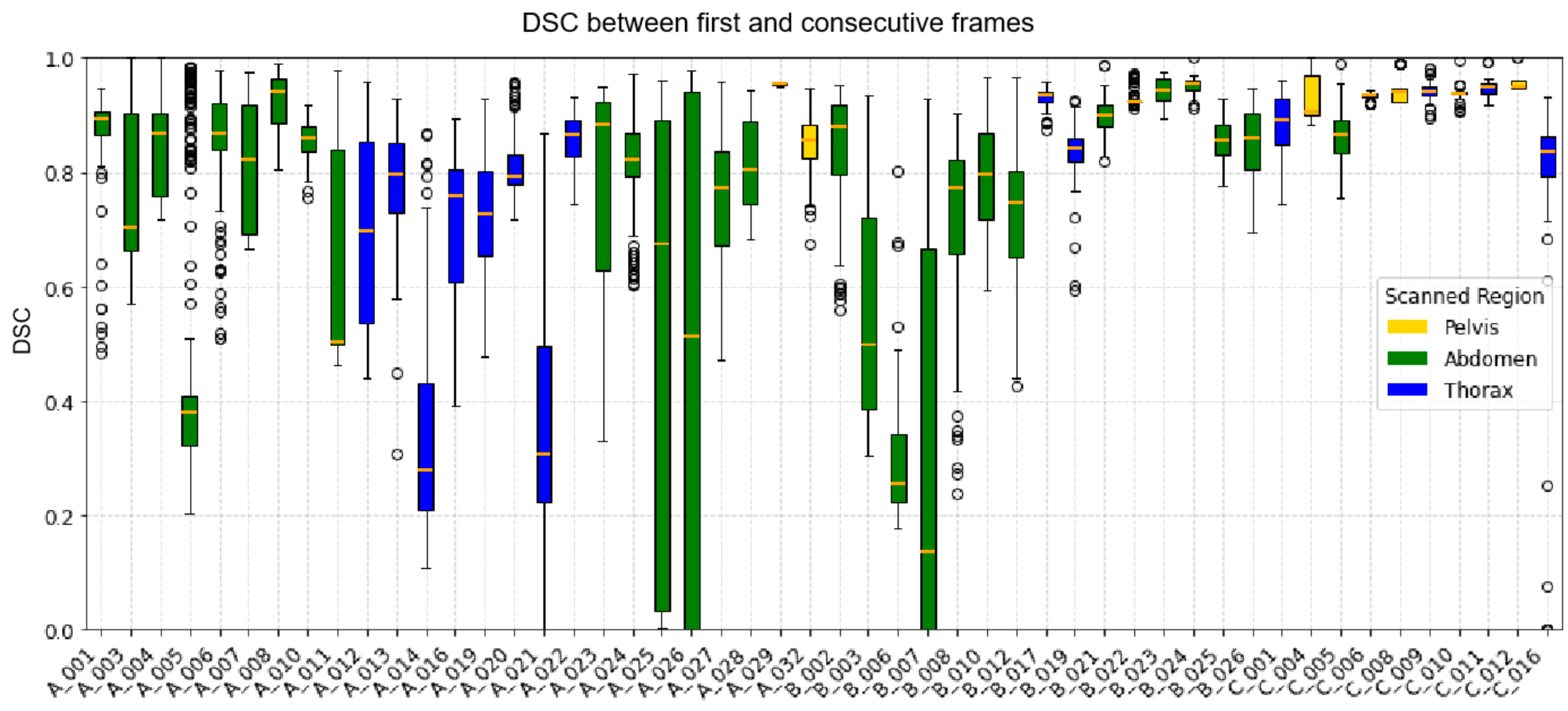

(a)

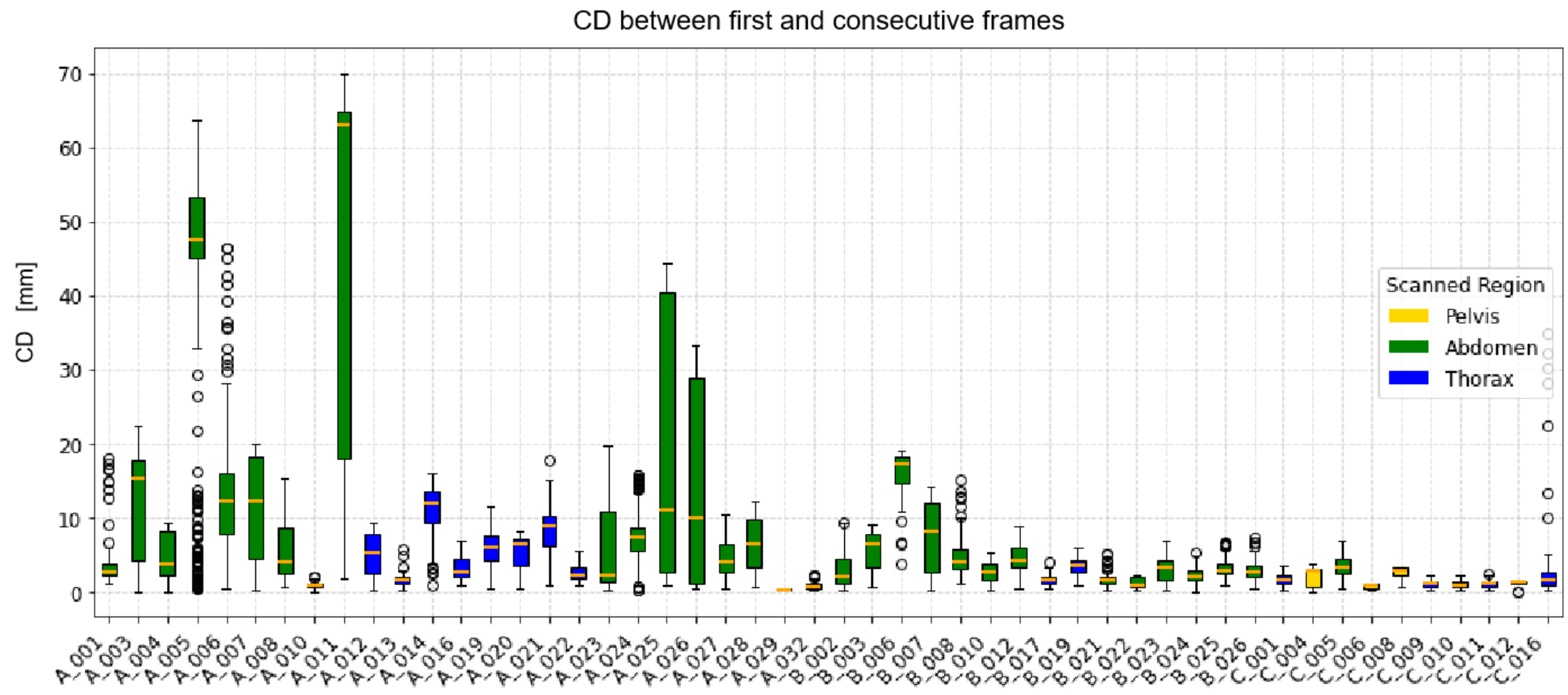

(b)

**Figure 4: The Dice similarity coefficient (DSC) (a) and Euclidean center distance (CD) (b) between the first and consecutive frames for the labeled 50 cases in the training set. The lower and upper box boundaries represent the 25th and 75th percentiles, respectively, the line inside the box represents the median, the whiskers extend to the 10th and 90th percentiles and the circles represent data falling outside the 10th and 90th percentiles.**

## 3. Data format and usage notes

### 3.1 Data structure and file formats

An overview of the folder structure for both the training and testing dataset is provided in Figure 5. Each patient has a unique alphanumeric name consisting of the data-providing center (A to F, X) and a three-digit patient ID. Each patient folder contains three JSON files and two sub-folders with cine MRI images and target segmentations, if available. The magnetic strength, acquisition frame rate, and scanned anatomical region are provided in the JSON files. The images subfolder contains all cine MRI frames assembled into a single MHA file named `<alphanumeric_name>_frames.mha`. If multiple MRIs have been acquired for the same patient, the image folder will contain additional files named `<alphanumeric_name>_frames<scan number>.mha`. The targets subfolder contains the label of the first cine MRI frame as `<alphanumeric_name>_first_label.mha` and the labels of the remaining frames assembled into `<alphanumeric_name>_labels.mha`. If multiple observers labeled the target, their labels are provided as `<alphanumeric_name>_labels<observer_ID>.mha.

The data is released under the CC BY-NC (Attribution-NonCommercial) license ((https://creativecommons.org/licenses/by-nc/4.0/deed.en). The training data is available on HuggingFace under the following link: https://doi.org/10.57967/hf/4539. It can be downloaded since March 15th, 2025. Preliminary testing and final testing datasets are only available for evaluation via submission to the grand-challenge platform: https://trackrad2025.grand-challenge.org/submission-instructions/. The testing set data will be uploaded to HuggingFace when the challenge is closed, except for center D where this is not allowed due to data protection regulations.

```
dataset/
|-- <patient>/
|   |-- field-strength.json
|   |-- frame-rate.json
|   |-- frame-rate<scan>.json -> additional scans (for some of the unlabeled patients)
|   |-- scanned-region.json
|   |-- scanned-region<scan>.json -> additional scans (for some of the unlabeled patients)
|   |-- images/
|   |   |-- <patient_id>_frames.mha
|   |   `-- <patient_id>_frames<scan>.mha -> additional scans (for some of the unlabeled patients)
|   `-- targets/
|       |-- <patient_id>_first_label.mha
|       |-- <patient_id>_labels.mha
|       `-- <patient_id>_labels<observer>.mha -> additional observers (for some of the labeled patients)
|-- D_001/ -> this is a labeled patient
|   |-- field-strength.json -> 0.35
|   |-- frame-rate.json -> 4.0
|   |-- scanned-region.json -> "thorax"
|   |-- images/
|   |   `-- D_001_frames.mha
|   `-- targets/
|       |-- D_001_first_label.mha
|       `-- D_001_labels.mha
|-- F_002/ -> this is an unlabeled patient
|   |-- field-strength.json -> 1.5 --> same for all scans of one patient
|   |-- frame-rate.json -> 1.65 -> frame rate of first scan
|   |-- frame-rate2.json -> 1.3 -> frame rate of second scan
|   |-- frame-rate3.json -> 1.65 -> frame rate of third scan
|   |-- scanned-region.json -> "abdomen" -> anatomic region of first scan
|   |-- scanned-region2.json -> "pelvis" -> anatomic region of second scan
|   |-- scanned-region3.json -> "abdomen" -> anatomic region of third scan
|   `-- images/
|      |-- F_002_frames.mha -> first scan
|      |-- F_002_frames2.mha -> second scan
|      `-- F_002_frames3.mha -> third scan from the same patient
|-- F_003/
`-- ...
```

**Figure 5: Folder structure of the TrackRAD2025 datasets.**

## 3.2 Usage notes

Except for the JSON files, all images and targets are provided in MHA format, which can be read and modified using the ITK open-source framework (https://itk.org/). SimpleITK (https://simpleitk.org/) provides a simplified interface to ITK for various languages, such as Python, R, Java, and C++. An example dataset, following the format of the labeled TrackRAD2025 dataset, as well as the recommended folder structure and the algorithm to read the image files, can be found at https://github.com/LMUK-RADONC-PHYS-RES/trackrad2025/. To view MHA images with a graphical user interface, 3DSlicer (https://www.slicer.org/), vv (https://github.com/open-vv/vv), and other open-source software for image processing can be used.

## 4. Discussion

Recent developments in MRIgRT offer the possibility of adapting the irradiation to changes in tumor position in real time during the treatment. 2D cine-MRI enables such real-time tumor visualization, but accurate tumor localization on all time-resolved frames is required to ensure the sparing of critical organs. The dataset may exhibit variability in image quality due to differences in clinical protocols across centers. This will require algorithms participating in TrackRAD2025 to be robust in multi-center settings, which is a realistic scenario. We expect TrackRAD2025 to impact the field of MRIgRT by providing the first public cine MRI dataset and benchmarking of real-time tumor localization algorithms based on a unified platform.

Determining the best-performing algorithm for tumor tracking on 2D cine MRI at MRI-linacs may benefit patients with motion-affected tumors by paving the way for MRI-guided free-breathing MLC tracking [29] instead of beam gating in combination with breath-holds, making treatments more efficient and cost-effective.

A limitation of the dataset is that it is restricted to 2D sagittal planes, which poses challenges for out-of-plane motion, particularly in the left-right direction. However, this single-plane approach reflects the current clinical standard. Future developments are expected to shift toward multi-2D acquisitions. With multi-plane cine MRI now feasible in clinical settings [11, 18], we plan to expand the dataset in a future edition of the challenge to include multi-plane cine MRI. This expansion may allow developers to explore models for tracking across multiple planes, ultimately enabling real-time 3D+t target tracking and motion management. Advanced 3D+t tracking could support enhanced MLC tracking capabilities, improving treatment conformality by dynamically adjusting the beam to the deforming target in real-time and incorporating dose-based optimization to maintain treatment goals despite anatomical changes [30].

Additionally, the dataset is limited to thoracic, abdominal, and pelvic tumors, which may not fully capture the diversity of tumor types and locations encountered in clinical radiotherapy. Expanding the dataset to include a broader range of tumor sites could enhance its applicability and support the development of more generalized motion management models. However, the selected anatomical locations should cover the majority of tumors affected by motion.

## 5. Conclusion

The TrackRAD2025 dataset provides a valuable resource for developing and evaluating real-time tumor localization algorithms for MRI-guided radiotherapy using clinical multi-centric data. This dataset fosters innovation in adaptive radiotherapy by enabling a standardized and fair comparison of real-time tumor tracking methods through the upcoming TrackRAD2025 challenge. The ability to accurately and efficiently localize tumors in real time has significant implications for improving treatment precision, reducing margins, and enhancing patient outcomes. We anticipate that the TrackRAD2025 dataset will support the advancement and clinical adoption of robust tumor-tracking algorithms, ultimately contributing to an improved standard of care in radiotherapy.

## Acknowledgments

This work was partly supported by Sichuan Science and Technology Program under Grant 2023YFH0079. The authors would like to thank the BZKF Lighthouse Image-Guidance in Local Therapies for supporting the data collection for this dataset.